\documentclass[conference]{IEEEtran}
\PassOptionsToPackage{protrusion, expansion}{microtype} 

\usepackage{tikz}
    \usetikzlibrary{positioning}
    
\usepackage{tgheros}
\usepackage{inconsolata}

\usepackage{listings}
    \definecolor{RED}{HTML}{c9171e}
    \definecolor{BLUE}{rgb}{.0, .2, .6}
    \definecolor{BLUEalt}{HTML}{1e50a2}

\usepackage{listings}						
\usepackage{amsmath, amsthm}
    
\usepackage{graphicx, xcolor} 
\usepackage{url}
\usepackage{adjustbox}
\usepackage{booktabs}
\usepackage{multirow}
\usepackage{caption}
\usepackage{subcaption}
\usepackage{enumitem}
\usepackage{lipsum}
\usepackage{setspace}
\usepackage{booktabs}
\usepackage{multirow}
\usepackage{cleveref}
\usepackage{listings}
\usepackage{algorithm, algpseudocode}
    \algrenewcommand{\alglinenumber}[1]{{\scriptsize\bfseries\ttfamily\color{RED}#1}}

\usepackage{footnote}
\makesavenoteenv{tabular}
\makesavenoteenv{table}
\usepackage{cite}

\IEEEoverridecommandlockouts

\usepackage[acronym]{glossaries}
\makeglossaries
 
\newglossaryentry{fclayer}{
    name={\texttt{fc}-layer},
    description={fully connected layer}
}
 
\newglossaryentry{szfull}{
    name=\textsc{SZ lossy compression},
    description={SZ lossy compression}
}

\newglossaryentry{deepsz}{
    name=\textsc{DeepSZ},
    description={DeepSZ}
}

\newglossaryentry{dataArray}{
    name=\textsc{data array},
    description={}
}

\newglossaryentry{indexArray}{
    name=\textsc{index array},
    description={}
}
 
\newacronym{gcd}{GCD}{Greatest Common Divisor}
\newacronym{lcm}{LCM}{Least Common Multiple}

\begin{document}

\title{LCFI: A Fault Injection Tool for Studying Lossy Compression Error Propagation in HPC Programs}

\author{\IEEEauthorblockN{Baodi Shan,\IEEEauthorrefmark{1}
Aabid Shamji,\IEEEauthorrefmark{2}
Jiannan Tian,\IEEEauthorrefmark{1}
Guanpeng Li,\IEEEauthorrefmark{2}
and Dingwen Tao\IEEEauthorrefmark{1}
\thanks{Corresponding author: Dingwen Tao, School of Electrical Engineering and Computer Science, Washington State University, Pullman, WA 99164, USA.}
}
\IEEEauthorblockA{\IEEEauthorrefmark{1}School of Electrical Engineering and Computer Science, Washington State University, WA, USA\\ lwshanbd@gmail.com, \{jiannan.tian, dingwen.tao\}@wsu.edu}
\IEEEauthorblockA{\IEEEauthorrefmark{2}Department of Computer Science, University of Iowa, IA, USA\\
\{aabid-shamji, guanpeng-li\}@uiowa.edu}}

\maketitle

\pagestyle{empty}

\begin{abstract}
Error-bounded lossy compression is becoming more and more important to today's extreme-scale HPC applications because of the ever-increasing volume of data generated because it has been widely used in in-situ visualization, data stream intensity reduction, storage reduction, I/O performance improvement, checkpoint/restart acceleration, memory footprint reduction, etc. 
Although many works have optimized ratio, quality, and performance for different error-bounded lossy compressors, there is none of the existing works attempting to systematically understand the impact of lossy compression errors on HPC application due to error propagation. 

In this paper, we propose and develop a lossy compression fault injection tool, called \textsc{LCFI}. To the best of our knowledge, this is the first fault injection tool that helps both lossy compressor developers and users to systematically and comprehensively understand the impact of lossy compression errors on HPC programs. The contributions of this work are threefold: (1) We propose an efficient approach to inject lossy compression errors according to a statistical analysis of compression errors for different state-of-the-art compressors. 
(2) We build a fault injector which is highly applicable, customizable, easy-to-use in generating top-down comprehensive results, and demonstrate the use of \textsc{LCFI}.
(3) We evaluate \textsc{LCFI} on four representative HPC benchmarks with different abstracted  fault models and make several observations about error propagation and their impacts on program outputs. 
\end{abstract}

\section{Introduction}
\label{sec:intro}

Today's HPC simulations and advanced instruments produce vast volumes of scientific data, which may cause many serious issues, including a huge storage burden  \cite{cesm-le-data,baker2014methodology,sz18,li2018optimizing}, I/O bottlenecks compared with fast stream processing \cite{cappello2019use}, and insufficient memory issues \cite{wu2018memory}. 
For example, the Hardware/Hybrid Accelerated Cosmology Code (HACC) \cite{hacc} (twice a finalist nomination for ACM's Gordon Bell Prize) can produce 20 petabytes of data to store when simulating up to 3.5 trillions of particles with 300 timesteps. 
Even considering a sustained bandwidth of 1 TB/s, the I/O time will still exceed 5 hours, which is prohibitive. Thus, the researchers generally output the data by decimation, that is, storing one snapshot every several timesteps in the simulation. This process definitely degrades the temporal constructiveness of the simulation and also loses valuable information for post-analysis. 

Another typical example is instrument data generated for materials science research. The advanced instruments (such as the Advanced Photon Source at Argonne) may produce the data with a super-high rate such as 500 GB/s (will increase by at least two orders of magnitude with the coming upgrades \cite{APSU}) so that thousands of discs are required to sustain the high data production rate if without compression support.

To mitigate the significant storage burden and I/O bottleneck, researchers have used many data compressors. Lossless compressors such as Gzip \cite{gzip}, Zstd \cite{zstd}, Blosc \cite{blosc}, and FPC \cite{fpc} suffer from low compression ratios (around 2:1 \cite{son2014data}) in reducing scientific data size because of the high randomness of ending mantissa bits in the floating-point representations \cite{Lindstrom2017Error}. Accordingly, error-bounded lossy compression has been treated as one of the best approaches to solve this big scientific data issue \cite{sz16,li2018optimizing}.  

Although existing error-bounded lossy compressors such as SZ \cite{sz16,sz17,sz18} and ZFP \cite{zfp} can strictly control the compression error of each data point, a significant gap still remains in understanding the impact of compression errors on program output. In other words, the propagation of compression errors in HPC programs has not been well studied and understood. 
Therefore, current lossy compression methods may lead to unacceptably inaccurate results for scientific discovery \cite{sasaki2015exploration,calhoun2019exploring,reza2019analyzing} based on the corrupted program output. 

Fault Injection (FI) is a widely used technique to evaluate the resilience of software applications to faults. While FI has been extensively used in general purpose applications, to the best of our knowledge, there does not exist a FI tool for lossy compression errors. The main challenges in developing such a fault injector remain in (1) designing a proper abstraction of compression fault model, and (2) integrating the fault model at the level where one can also conduct program-level error propagation analysis.
Our contributions are listed as follows.
\begin{itemize}
    \item We propose a systematic approach for efficient lossy compression fault injection to help compressor developers and users to understand the impact of compression error on their interest in HPC applications.
    \item We build a fault injector (called \textsc{LCFI}) to inject lossy compression errors into any given HPC program. The tool is highly applicable, customizable, easy-to-use, and able to generate top-down comprehensive results. We also demonstrate the use of \textsc{LCFI} using an example program.
    \item We evaluate \textsc{LCFI} on four representative HPC benchmark programs with different abstracted lossy compression fault models to understand how different compressors affect those programs' outputs. Experimental results provide several important insights for users to understand how to strategically use lossy compression in order to avoid corrupting program output.
\end{itemize}

The rest of the paper is organized as follows.
In Section~\ref{sec:background}, we discuss the background and our research motivation. 
In Section~\ref{sec:model}, we discuss our fault model for lossy compression error.
In Section~\ref{sec:design}, we present the design and implementation details of our FI tool \textsc{LCFI}.  
In Section~\ref{sec:usage}, we describe the use of \textsc{LCFI} in detail. 
In Section~\ref{sec:evaluation}, we present our evaluation results.
In Section~\ref{sec:conclusion}, we conclude and discuss future work.

\section{Background and Motivation}
\label{sec:background}

\subsection{Error-bounded Lossy Compression for HPC Data}
Data compression has been studied for decades. There are two main categories: lossless compression and lossy compression. Lossless compressors such as FPZIP~\cite{lindstrom2006fast} and FPC~\cite{fpc} can only provide limited compression ratios (typically up to 2:1 for most scientific data) due to the significant randomness of the ending mantissa bits~\cite{son2014data}.

Lossy compression, on the other hand, can compress data with little information loss in the reconstructed data.
Compared to lossless compression, lossy compression can provide a much higher compression ratio while still maintaining useful information for scientific discoveries. 
Different lossy compressors can provide different compression modes, such as error-bounded mode and fixed-rate mode. 
Error-bounded mode requires users to set an error bound, such as absolute error bound and point-wise relative error bound. The compressor ensures the differences between the original data and the reconstructed data do not exceed the user-set error bound.
Fixed-rate mode means that users can set a target bitrate, and the compressor guarantees the actual bitrate of the compressed data to be lower than the user-set value.
In this work, we mostly focus on the error-bound mode and leave the fixed-rate mode for the future work. 

In recent years, a new generation of lossy compressors for HPC data has been proposed and developed, such as SZ~\cite{sz16, sz17, sz18} and ZFP~\cite{zfp}.
Unlike traditional lossy compressors such as JPEG \cite{wallace1992jpeg} ,which is designed for images (in integers), SZ and ZFP are designed to compress floating-point and integer HPC data and can provide a strict error-controlling scheme based on user's requirements.
SZ is a representative prediction-based error-bounded lossy compressor. SZ has three main steps: (1) predicts each data point's value based on its neighboring points by using an adaptive, best-fit prediction method; (2) quantizes the difference between the real value and predicted value based on the user-set error bound; and (3) applies a customized Huffman coding and lossless compression to achieve a higher compression ratio.
ZFP is a representative transform-based error-bounded lossy compressor for floating-point and integer data.  ZFP splits the whole data set into many small blocks with an edge size of 4 along each dimension and compresses the data in each block separately in four main steps:  (1) alignment of exponent, (2) orthogonal transform, (3) fixed-point integer conversion, and (4) bit-plane-based embedded coding. 
For more details, we refer readers to \cite{sz17} and \cite{zfp} for SZ and ZFP, respectively.

\subsection{LLFI}
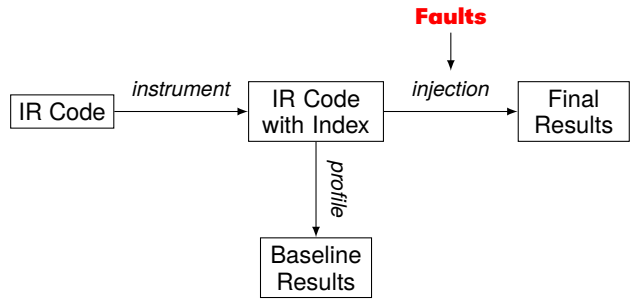
\begin{figure}[h]
\centering\small
\sffamily
\begin{tikzpicture}
\node[draw] (ir-code) {IR Code};
\node[draw, text width=5em, align=center, right=.7 in of ir-code] (ir-code-idx) {IR Code with Index};
\node[draw, text width=4em, align=center, right=.7in of ir-code-idx] (final-res) {Final Results};
\node[draw, text width=4em, align=center, below=.5in of ir-code-idx] (baseline-res) {Baseline Results};

\draw[-latex] 
    (ir-code) 
    to[] node[above, font=\itshape\footnotesize\strut] {instrument} (ir-code-idx);
\draw[-latex] 
    (ir-code-idx) 
    to[] node (inject) [above, font=\itshape\footnotesize\strut] 
    {injection}
    (final-res);
\draw[-latex]
    (ir-code-idx)
    to[] node[midway, above, sloped, font=\itshape\footnotesize\strut]
    {profile}
    (baseline-res);
\node[above=.2in of inject] (fault) {\color{red}\small\fontfamily{ugq}\selectfont Faults};
\draw[-latex]
    (fault)
    --
    (inject.north);
\end{tikzpicture}
\caption{Overview of LLFI workflow.}
\label{fig:design1}
\end{figure}

LLFI\cite{LLFI-QRS} is an LLVM based FI tool that injects faults into the LLVM IR of the application source code. There are three core parts in LLFI: \textit{Instrument}, \textit{Profile}, and \textit{Injection} as shown in Figure \ref{fig:design1}.

In general, the instrument part takes an IR file as input and generates IR files with instrumented profiling and fault injection function calls. The profile part takes a profiling executable, executes it, and generates the baseline results. Using these results, users can determine whether the fault has influenced the execution of the program. The injection part will inject a fault set in the \textit{input.yaml} to the program. After this step, the final results are generated including \textit{program output file}, \textit{trace file} and \textit{fault-injection file}.

\subsection{Research Motivation}

Existing lossy compressors mainly focus on optimizing from three aspects: compression ratio (i.e., storage reduction ratio), and compression speed (a.k.a., throughput), and reconstructed data quality based on statistical metrics such as PSNR (peak signal-to-noise ratio) and SSIM (structural similarity index measure). However, only few works \cite{tao2018improving,reza2019analyzing,evans2020jpeg} have studied the impact of compression error on HPC applications and none of them have systematically studied how compression errors propagate in any HPC program. This is because unlike traditional resilience and fault tolerance community that has many fault injection tools (such as PinFI \cite{PINFI}, LLFI \cite{LLFI-QRS}, and TensorFI \cite{li2018tensorfi}) to investigate how software applications are resilient to hardware errors, the HPC community is missing an efficient fault injection tool for lossy compression error, which can help compressor developers and users to understand the compression error impact on specific HPC programs. This motivates us to develop such a tool in this work. 
\section{Lossy Compression Fault Model} 
\label{sec:model}

Unlike lossless compression, lossy compression cannot precisely recover numerical data bit by bit. However, lossy compressed data are acceptable in many use cases (such as storage reduction, in situ visualization, and checkpoint/restart \cite{cappello2019use}) for HPC applications. This is because HPC/scientific data itself tends to involve many error terms. 
Taking experimental and observational data as an example, finite precision measurements and intrinsic measurement noise make an impact on the data accuracy. 
On the other hand, round-off, truncation, and model errors that appear in numerical simulations also have limited precision. Thus, using lossy compression techniques to approximate floating-point data is acceptable and even one of the most promising solutions for solving the big scientific data issue \cite{calhoun2019exploring, poppick2020statistical, li2018data}. 

We propose to simulate compression errors instead of performing actual compression and decompression for FI because current state-of-the-art lossy compressors such as SZ and ZFP can only provide the throughputs of hundreds of megabytes per second. Taking into account the following two reasons, the approach of actual compression and decompression would introduce very high runtime overheads:
(1) existing lossy compressors have a large design space including compression algorithms (such as SZ \cite{sz16,sz17,sz18}, ZFP \cite{zfp}, FPZIP \cite{fpzip}, MGARD \cite{mgard}, TTHRESH \cite{tthresh}, VAPOR \cite{vapor}, etc.) and their diverse configurations (e.g., error-bound mode and value); and
(2) in order to obtain a reasonable coverage for diverse HPC programs, a large amount of FI locations need to be considered.
As a result, the approach of actual compression and decompression for FI is very inefficient. Therefore, we choose to simulate the compression errors in our FI tool. 

\begin{figure}
    \centering
	\begin{subfigure}{\linewidth}\centering
	    \includegraphics[width=\linewidth]{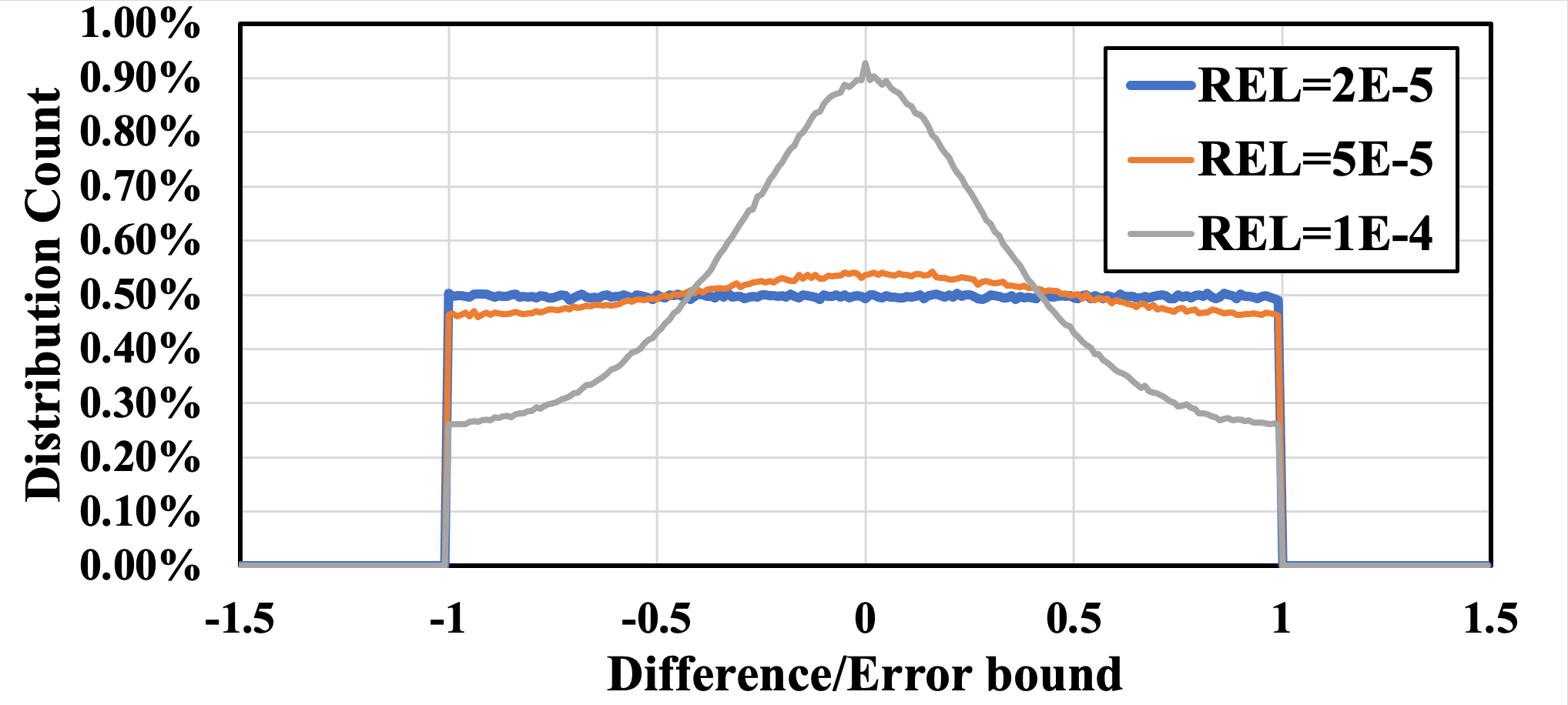}
	    \caption{\footnotesize SZ}\label{fig:Nyx_Temp_Error_Dis}
	\end{subfigure}
	\begin{subfigure}{\linewidth}\centering
	    \includegraphics[width=\linewidth]{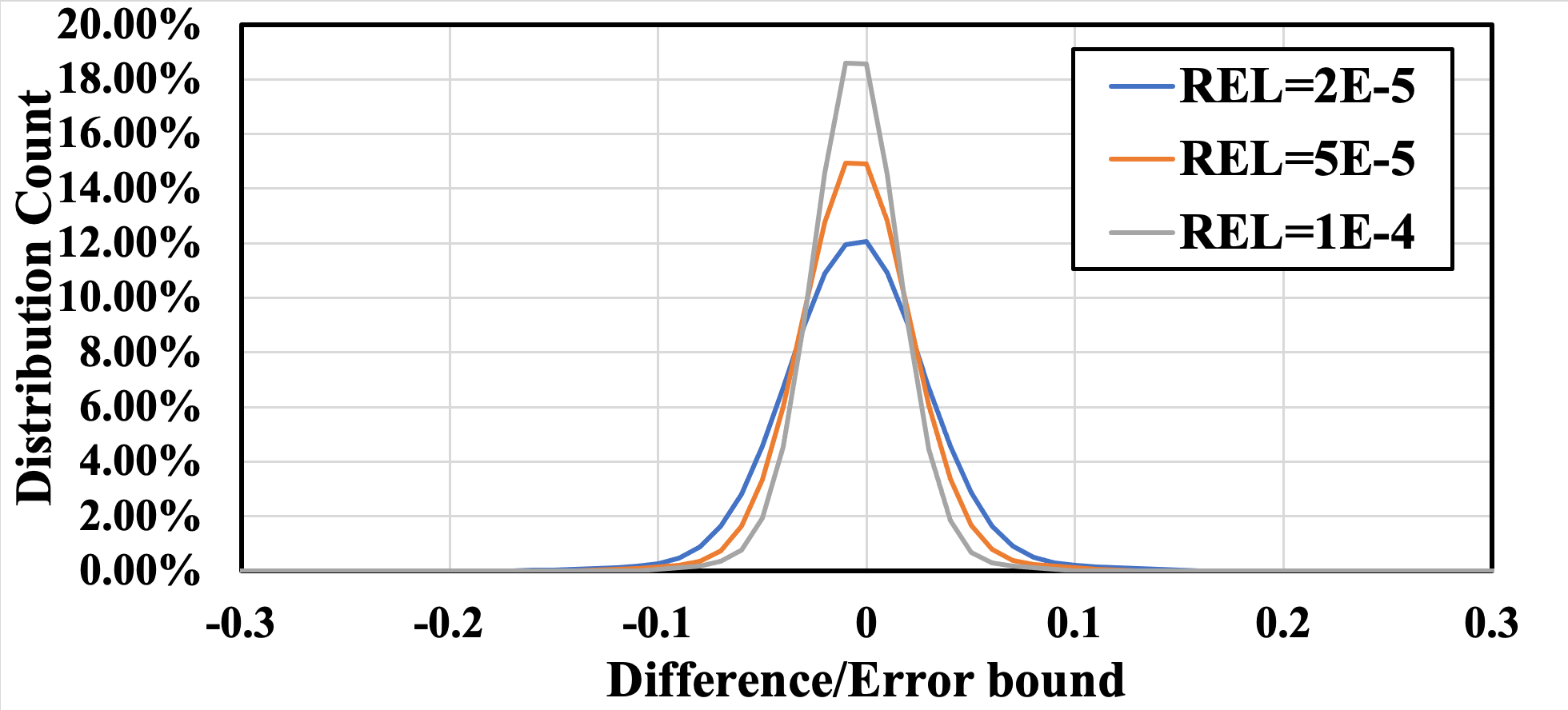}
	    \caption{\footnotesize ZFP}\label{fig:Nyx_Vx_Error_Dis}
	\end{subfigure}
\caption{Example error distributions of SZ and ZFP on a typical variable (temperature) in Nyx dataset with different error bounds. The bin size for histogramming is $0.01\cdot eb$.}
    \label{fig:fig-31-1}
\end{figure}

To simulate the compression errors, we have to understand the fault model for a specific compression algorithm.
For example, Figure~\ref{fig:fig-31-1} illustrates an example error distribution when compressing and decompressing a typical variable in Nyx cosmology data.
It clearly shows that there exists an identifiable error distribution with different compression configurations of the SZ and ZFP compression algorithms. 
In fact, Lindstrom \cite{Lindstrom2017Error} studied errors distributions of lossy floating-point compressors in a statistical way. The work concludes that lossy compression error distributions depend on their adopted quantization techniques. Specifically, lossy compressors adopted uniform scalar quantization such as SZ \cite{sz16,sz17,sz18}, SQ \cite{iverson2012fast}, and LZ4A \cite{kunkel2017toward} tend to generate uniformly distributed errors, while transform-based lossy compressors such as ZFP \cite{zfp}, VAPOR \cite{vapor}, and TTHRESH \cite{tthresh} produce error distributions that are close to normal (a.k.a., Gaussian). 
Inspired by this work, we mainly focus on these two fault models (i.e., uniform and normal distributions) in this study; however, it is worth noting that \textsc{LCFI} is extensible with any given error distribution (will be described later).

\section{Design and Implementation}
\label{sec:design}

\textsc{LCFI}\footnote{\textsc{LCFI} is publicly available at \url{https://github.com/LCFI/LCFI}.} is an extension of LLFI\cite{LLFI-QRS}. In this Section, we first discuss our design goals and assumptions for \textsc{LCFI}. We then present the improvements and features of \textsc{LCFI}. Finally, we present our implementation details.

\subsection{Design Goals and Assumptions}

In general, we have four design goals for \textsc{LCFI} as follow:

\begin{itemize}
    \item \textbf{Applicability:} We aim to create a tool that is simple and easy to use, that users can exploit even without knowing a lot about error-bounded lossy compression. With a simple program written in C/C++, users should be able to easily inject a fault to  \textit{a specific variable} at \textit{a specific location}. For example, if the target variable is located in a for-loop, the user can inject faults in a specific iteration of this for-loop, which is necessary to change an array's value.
    \item \textbf{Customizable:}  Given that there are a large number of error distributions in lossy compression (considering future newly designed compressors), it is not feasible to provide a tailored tool for all distributions. We provide a template to users to allow them customize their own error distributions.
    \item \textbf{Easy-to-Use:} We aim to provide users a simple installation process that does not require editing several setup files. To install \textsc{LCFI}, users only need to edit just one or two YAML-files and run a few commands (e.g., no more than four) to get the injection results. Moreover, \textsc{LCFI} should not require an understanding of how the compiler works or the ability to read IR files. 
    \item \textbf{Top-Down Comprehensive Result:} We aim to make the injector provide both high-level and underlying results (such as registers' value). Users can choose to revise the output file or trace the error propagation to potentially find Benign Faults \cite{li2018modeling} (will be discussed in Section \ref{sec:evaluation}).
\end{itemize}

Additionally, we make the following assumptions about the faults injected by \textsc{LCFI}:

\begin{itemize}
    \item Faults can only be injected into variables that are on the right of the equal sign due to the nature of LLVM. Changing a variable on the left of the assignment can be achieved by changing all variables on the right of the assignment. 
    \item\label{sec:limit} Faults cannot be injected to the variables located in the \textit{main} function. This is because most of the faults in the \textit{main} functions will cause the program to crash, which will make injection meaningless.
    \item Because LLFI does not support OpenMP, one can only run \textsc{LCFI} on serial programs without multiple threads. In the future, with LLFI-GPU developed, we will further design an OpenMP and CUDA version of \textsc{LCFI}.
\end{itemize}

\subsection{Design of LCFI}

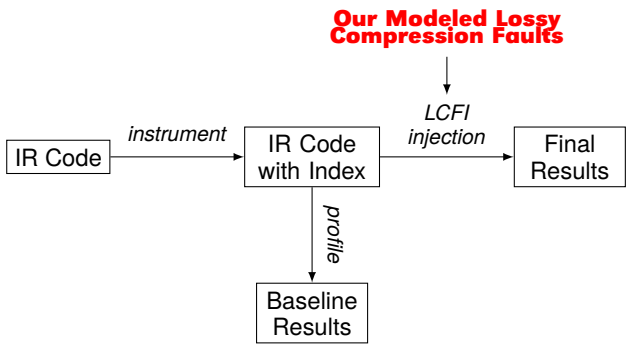
\begin{figure}[h]
\centering\small
\sffamily
\begin{tikzpicture}
\node[draw] (ir-code) {IR Code};
\node[draw, text width=5em, align=center, right=.7 in of ir-code] (ir-code-idx) {IR Code with Index};
\node[draw, text width=4em, align=center, right=.7in of ir-code-idx] (final-res) {Final Results};
\node[draw, text width=4em, align=center, below=.5in of ir-code-idx] (baseline-res) {Baseline Results};

\draw[-latex] 
    (ir-code) 
    to[] node[above, font=\itshape\footnotesize\strut] {instrument} (ir-code-idx);
\draw[-latex] 
    (ir-code-idx) 
    to[] node (inject) [text width=4em, align=center, above, font=\itshape\footnotesize\strut] 
    {LCFI\\injection}
    (final-res);
\draw[-latex]
    (ir-code-idx)
    to[] node[midway, above, sloped, font=\itshape\footnotesize\strut]
    {profile}
    (baseline-res);
\node[above=.2in of inject, text width=10em, align=center] (fault) {\color{red}\small\fontfamily{ugq}\selectfont Our Modeled Lossy\\[-.8ex]Compression Faults};
\draw[-latex]
    (fault)
    --
    (inject.north);
\end{tikzpicture}
\caption{Overview of our proposed \textsc{LCFI} workflow.}
\label{fig:design2}
\end{figure}

Unlike LLFI that focuses on the impact caused by different software faults and hardware faults, \textsc{LCFI} focuses on how different lossy compression errors impact the running of different programs. Thus, to build LCFI, we modify the way LLFI injects faults and faults themselves. The core design of \textsc{LCFI} is shown in Figure \ref{fig:design2}. 

We propose the following designs in \textsc{LCFI} to satisfy the previously described goals. More details will be shown later in Section \ref{sec:model}.
\begin{itemize}
    \item \textbf{Applicability:} We provide several YAML files that users can edit. In these YAML files, users can easily select the variable where they want to inject the fault and select what kind of fault model they want to inject. Users are not required to understand how lossy compression works but can still get results directly.
    \item \textbf{Customizable:} Unlike LLFI's complex step of customizing faults, we provide a template for the distribution of lossy compression errors. To custom faults or error distributions, users just need to simply edit this template and recompile the code.
    \item \textbf{Ease-to-Use:} By using the Python scripts that we provide, \textsc{LCFI} can automatically find the location of specific variables in the IR file. Users can use the scripts to notify the injector what index it should target. Thus, users do not need to understand a complicated IR file to use \textsc{LCFI}. 
    \item \textbf{Top-Down Comprehensive Result:} \textsc{LCFI} generates both high-level and underlying results such as standard output files and IR-level results. Users can use both results to perform program-level error propagation analysis.
\end{itemize}

\subsection{LCFI Features}

\textsc{LCFI} improves the functionality of LLFI by introducing the following new features:
\begin{itemize}
    \item \textbf{Multi-location} Unlike LLFI that can inject a fault to only one specific location, \textsc{LCFI} allows to inject a fault at \textit{any given location} and at \textit{any given time}.
    \item \textbf{For-loop Injection} For HPC programs, for-loop is one of the most frequently used loops. For LCFI, we design an interface to set the loop number so that users can inject faults at specific iterations during the for-loop execution. This is imperative if the user wants to inject the fault into an array.
    \item \textbf{Custom Distribution} We optimize the current LLFI interface to allow users to easily \textit{customize} their own lossy compression errors.
\end{itemize}

\subsection{Implementation Details}

Similar to LLFI, \textsc{LCFI} is implemented using the LLVM-Pass (in C/C++) and Python. We split the LLVM-Pass code and Python code into three modules as follows:

\begin{itemize}
    \item \textbf{LLVM-Pass Core} is the main module that controls the underlying execution of the target program. It also provides the functionality to trace the execution and insert runtime code.
    \item \textbf{Runtime Lib} module consists of different fault implementations and determines which variables need to be injected.
    \item \textbf{Tools} module consists of some useful tools for users to analyze the results from LCFI. It includes Trace\_To\_Dot, Trace\_Union and Trace\_Diff.  
\end{itemize}

\textsc{LCFI} results consist of four main outputs as follows: 

\begin{itemize}
    \item \textbf{Baseline:} This output comes from the origin program, which includes \textit{golden\_std\_output}, \textit{llfi.stat.trace.prof.txt} and an output file. \textit{golden\_std\_output} is the standard output of the origin program. The \textit{llfi.stat.trace.prof.txt} records the value changes of every register. 
    \item \textbf{Program Output:} This output comes from the execution of the  program with injected faults. If the program does not generate an output file, this part will be empty.
    \item \textbf{Error Output:} If the program with injected faults crashes, the log will be stored in this output file.
    \item \textbf{Standard Output:} This file records the execution of the program with injected faults.
    \item \textbf{LLFI Stat Output:} This file records the value change of every register. If faults have successfully been injected into the program, the injection log will also be stored.
\end{itemize}

\section{Usage Model}
\label{sec:usage}
In this section, we will demonstrate how to customize a distribution of lossy compression errors in \textsc{LCFI} and how to inject the fault into a program written in C/C++. The example C code is shown in \cref{lst:animate}.

\begin{lstlisting}[caption={An example C code for demonstration.},label={lst:animate}]
#include<stdio.h>

double process(double n[])
{
	double ans=0;
	for(int i=0;i<3;i++)
	{
		ans=n[i]; 
		printf("n[%d]: %lf\n",i,ans);
	}
	return ans;
}

int main(){
	double n[3];
	freopen("in.txt","r",stdin); 
	freopen("output.txt","w",stdout);
  	scanf("%lf %lf %lf",&n[0],&n[1],&n[2]);
	double ans;
	ans=process(n);
	printf("++++++++++++++++++++++++\n");
	ans=process(n);
	printf("++++++++++++++++++++++++\n");
	ans=process(n);

	fclose(stdin);
	fclose(stdout);
	return 0;
}

\end{lstlisting}

In the sample code, the \textit{main} function calls the \textit{process} function three times. The \textit{process} function contains a for-loop that will be executed three times. In each for-loop, the program will print the value in the \textit{n} array.

\begin{lstlisting}[keywordstyle={\color{black}}, caption={Example configuration file in YAML format.},label={lst:yaml1}]
variable_num: 1
loop_num: 3
fi_type: Nor5(LCFI)
option:
  -
   function_name: process
   variable_name: n
   variable_init: true
   variable_location: 1
   in_arr: true
   in_loop: true
\end{lstlisting}

After compiling and instrumenting the C code, we will get three IR files. Let us take a look at the \textit{demo-lcfi\_index.ll} first. Listing \ref{lst:code2} shows its part related to \textit{process} function. In this file, every IR instruction is given an index so that the injector can recognize different instructions in the next step. We target to inject compression errors on the variable $n$ (Line 8 in the example code). To do so, we set \textit{variable\_name} as \textit{n} and \textit{function\_name} as the \textit{process} in the configurable YAML file, as shown in Listing \ref{lst:yaml1}. The variable \textit{n} first appears in Line 8, so we set \textit{variable\_location} as \textit{1}. Because \textit{n} is in a for-loop and is an array, we set the \textit{in\_arr} and \textit{in\_loop} to \textit{true}. In particular, as we target to inject faults in the 3rd loop, we set \textit{loop\_num} as \textit{3}. Running the python script \textit{setinput.py} generates \textit{input.yaml}.

\begin{lstlisting}[basicstyle=\ttfamily\scriptsize, keywordstyle={\color{black}}, caption={Details of demo-lcfi\_index.ll.},label={lst:code2}]
define double @process(double* %n) #0 {
  %1 = alloca double*, align 8, !llfi_index !1
  %ans = alloca double, align 8, !llfi_index !2
  %i = alloca i32, align 4, !llfi_index !3
  store double* %n, double** %1, align 8, !llfi_index !4
  store double 0.000000e+00, double* %ans, align 8, !llfi_index !5
  store i32 0, i32* %i, align 4, !llfi_index !6
  br label %2, !llfi_index !7

; <label>:2                                       ; preds = %14, %0
  %3 = load i32* %i, align 4, !llfi_index !8
  %4 = icmp slt i32 %3, 3, !llfi_index !9
  br i1 %4, label %5, label %17, !llfi_index !10

; <label>:5                                       ; preds = %2
  %6 = load i32* %i, align 4, !llfi_index !11
  %7 = sext i32 %6 to i64, !llfi_index !12
  %8 = load double** %1, align 8, !llfi_index !13
  %9 = getelementptr inbounds double* %8, i64 %7, !llfi_index !14
  %10 = load double* %9, align 8, !llfi_index !15
  store double %10, double* %ans, align 8, !llfi_index !16
  %11 = load i32* %i, align 4, !llfi_index !17
  %12 = load double* %ans, align 8, !llfi_index !18
  %13 = call i32 (i8*, ...)* @printf(i8* getelementptr inbounds ([12 x i8]* @.str, i32 0, i32 0), i32 %11, double %12), !llfi_index !19
  br label %14, !llfi_index !20

; <label>:14                                      ; preds = %5
  %15 = load i32* %i, align 4, !llfi_index !21
  %16 = add nsw i32 %15, 1, !llfi_index !22
  store i32 %16, i32* %i, align 4, !llfi_index !23
  br label %2, !llfi_index !24

; <label>:17                                      ; preds = %2
  %18 = load double* %ans, align 8, !llfi_index !25
  ret double %18, !llfi_index !26
}
\end{lstlisting}

Then, let us take a look at \textit{demo-lcfi\_profiling.ll}\footnote{This file is generated when trace option is set to \texttt{true}.} and \textit{demo-lcfi\_fi.ll}. Both files include some instructions that are used for printing trace information and fault injection. 
There are some instructions of which trace information is not added in front because these instructions do not return any registers. This kind of instructions always uses the same \textit{store} opcode because the \textit{store} instruction only stores some value in a specific register but does not return any registers. That is why assuming that users cannot change the variable on the left of the assignment symbol, as presented in Section \ref{sec:limit}. 

The next step is profiling and injection. After that, \textit{demo-profiling.ll} and \textit{demo-fi.ll} will be compiled to executable files. Then, we can get the results of the baseline program and program with injected faults by executing these executable files. If turning on the trace switch, we can also get trace files for baseline run and run with injected compression faults, similar to Listing \ref{lst:trace1} and \ref{lst:trace2}. We can use the \textit{trace-diff} command to analyze the error propagation in terms of program execution. As shown in the listings, the values of index-18 are different, which means \textit{ans} has been impacted by the compression errors injected to the variable \textit{n} with the index of 15.

\begin{lstlisting}[caption={Trace of Profile.},label={lst:trace1},
basicstyle=\ttfamily\footnotesize]
ID: 15   OPCode: load   Value: 4010000000000000
ID: 16   OPCode: store  Value: 00000000
ID: 17   OPCode: load   Value: 00000000
ID: 18   OPCode: load   Value: 4010000000000000
ID: 19   OPCode: call   Value: 0000000f
ID: 21   OPCode: load   Value: 00000000
ID: 22   OPCode: add    Value: 00000001
ID: 8    OPCode: load   Value: 00000001
\end{lstlisting}
\begin{lstlisting}[caption={Trace of Injected Fault.},label={lst:trace2}]
ID: 15   OPCode: load   Value: 4010000000000000
ID: 16   OPCode: store  Value: 00000000
ID: 17   OPCode: load   Value: 00000000
ID: 18   OPCode: load   Value: 4014e8d25119f5e3
ID: 21   OPCode: load   Value: 00000000
ID: 22   OPCode: add    Value: 00000001
ID: 8    OPCode: load   Value: 00000001
\end{lstlisting}

\begin{lstlisting}[caption={Details of output results.},label={lst:results}]
n[0]: 4.000000           n[0]: 4.000000
n[1]: 3.000000           n[1]: 3.000000
n[2]: 3.000000           n[2]: 3.000000
++++++++++++++++++++++++ ++++++++++++++++++++++
n[0]: 4.000000           n[0]: 4.000000
n[1]: 3.000000           n[1]: 3.000000
n[2]: 3.000000           n[2]: 3.000000
++++++++++++++++++++++++ ++++++++++++++++++++++
n[0]: 4.000000           n[0]: 2.699687
n[1]: 3.000000           n[1]: 3.253787
n[2]: 3.000000           n[2]: 4.396792
\end{lstlisting}

Finally, we can get the outputs of the baseline program and the program with injected faults, as shown in Figure \ref{lst:results}. The values in the third loop are all different from the baseline.
\section{Evaluation}
\label{sec:evaluation}

In this section, we use different compression fault models (i.e., error distributions) to inject faults into several representative HPC programs. In these programs, we select some typical variables for injection where lossy compression is needed. The names of programs and selected variables are shown in Table \ref{table:set}. The goal of our experiment is to demonstrate that \textsc{LCFI} has the ability to inject various compression faults with different error distributions into different program locations.

\begin{table*}[]\centering\sffamily
\footnotesize
\resizebox{0.75\textwidth}{!}{
\begin{tabular}{@{} *{7}{l} @{}}
\toprule
\bfseries Benchmark &
\bfseries Index &
\bfseries Variable Name &
\bfseries Data Type &
\bfseries In Array? &
\bfseries In For-Loop? &
\bfseries Loop Num. \\ 
\midrule
{HPCCG \cite{hpccg}} &
  {1469} &
  {x} &
  {Double} &
  {True} &
  {True} &
  {1, 5} 
\\ 
{Black-Scholes\cite{bs}} &
  {40} &
  {xNPrimeofX} &
  {Float} &
  {False} &
  {False} &
  {NaN} 
\\ 
{XSBench\cite{xsbench}} &
  {271} &
  {conc} &
  {Double} &
  {False} &
  {True} &
  {2} 
\\ 
{NPB-MG\cite{npbmg}} &
  {6326} &
  {a1} &
  {Double} &
  {False} &
  {False} &
  {NaN}  
\\ 
  \bottomrule
\end{tabular}
}
\caption{Configurations of tested benchmarks and targeted variables.}
\label{table:set}
\end{table*}

\subsection{Experimental Configuration}

Lossy compression is used for data reduction in HPC applications, thus, we select representative variables with  relatively large sizes for fault injection in the core function, as shown in Table \ref{table:set}.

\begin{itemize}
    \item \textbf{Programs:} We use the benchmarks provided by \cite{palazzi2019tale}, which are very popular HPC benchmarks.
    \item \textbf{Index and Variable Name:} In the IR format file, a specific llfi-index means a specific variable and its location. Using the index, we can determine the injection location.
    \item \textbf{In Loop or Array?:} The information of this attribute is discussed in Section \ref{sec:usage}.
    \item \textbf{Fault Type:} We use four types of fault models which are the combinations of two typical error-bound modes (absolute error and relative error) and two error distributions (uniform distribution and normal distribution).
\end{itemize}

\subsection{Evaluation Results}
\label{subsec:res}

\subsubsection{HPCCG}
\label{subsubsec:hpcg}

HPCCG is a simple conjugate gradient benchmark code for a 3D chimney domain. We test the variable \textit{x} in the \textit{waxpby} function. 
We observe that even injecting compression faults on the same variable, different error distributions or locations may lead to different program outputs. 

Table \ref{table:res1} shows the results when injecting faults in the first loop. 
We observe that every program with faults injected can still converge, but programs injected with absolute errors converge much slower than those with relative errors.

\begin{table*}[]\centering\sffamily
\footnotesize
\begin{tabular}{ @{} *{11}{l} @{}}
\toprule
    {\bfseries Fault Type}          
&   \multicolumn{5}{l}{Relative + Uniform} 
&   \multicolumn{5}{l}{Relative + Normal} 
\\ 
\cmidrule(r){1-1}\cmidrule(lr){2-6}\cmidrule(l){7-11}
    {\bfseries Error Bound}          
&   \multicolumn{5}{l}{1\%, 5\%, 10\%, 50\%, 100\%}       
&   \multicolumn{5}{l}{1\%, 5\%, 10\%, 50\%, 100\%}
\\ 
    {\bfseries Converge Iter.} 
&   \multicolumn{5}{l}{99 (same as baseline)}       
&   \multicolumn{5}{l}{99 (same as baseline)}      
\\ 
\bottomrule
\\
\toprule
    {\bfseries Fault Type}          
&   \multicolumn{5}{l}{Absolute + Uniform} 
&   \multicolumn{5}{l}{Absolute + Normal}
\\
\cmidrule(r){1-1}\cmidrule(lr){2-6}\cmidrule(l){7-11}
  {\bfseries Error Bound} &
  {0.01} &
  {0.05} &
  {0.1} &
  {0.5} &
  {1} &
  {0.01} &
  {0.05} &
  {0.1} &
  {0.5} &
  {1} 
\\ 
  {\bfseries Converge Iter.} &
  {103} &
  {103} &
  {104} &
  {105} &
  {105} &
  {103} &
  {104} &
  {105} &
  {105} &
  {105} \\ 
\bottomrule
\end{tabular}
\caption{Results of the first loop for \textit{x} in \textit{waxpby} function.}
\vspace{-4mm}
\label{table:res1}
\end{table*}

\begin{figure}
\centering
\includegraphics[width=0.4\paperwidth]{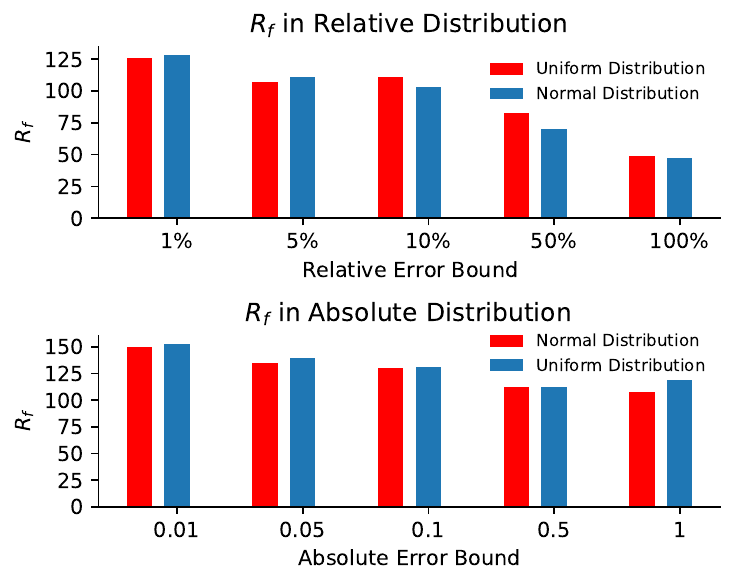}
\vspace{-2mm}
\caption{Negative logarithm of final residual  generated by different injected programs.}
\label{fig:res2}
\end{figure}

Moreover, when we inject faults on variable \textit{x} in the fifth loop, none of the programs is able to converge within 150 iterations (i.e., the maximum number of iterations set by the program in default). The results are shown in Figure \ref{fig:res2}. In order to better illustrate the \textit{final residual} of the program after 150 iterations, we compute a new metric $R_f$ as:
$$R_f=-log(f), f=final \text{ }residual.$$

\subsubsection{Black-Scholes}
\begin{figure}[H]
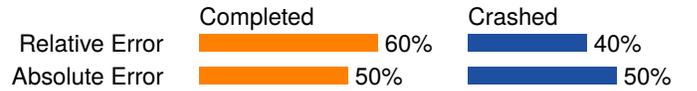

\centering\footnotesize\sffamily
\vspace{-2mm}
\resizebox{\linewidth}{!}{%
\begin{tabular}{@{}rll@{}}
            &   Completed       &   Crashed       \\
Relative Error &   {\color{orange}\rule{60pt}{6pt}} 60\%  &   {\color{BLUEalt}\rule{40pt}{6pt}} 40\%    \\[.5ex]
Absolute Error &   {\color{orange}\rule{50pt}{6pt}} 50\%  &   {\color{BLUEalt}\rule{50pt}{6pt}} 50\%    \\
\end{tabular}
}
\caption{Percentage of crashed and completed-but-corrupted runs.}
\vspace{-2mm}
\label{fig:res3}
\end{figure}

Black-Scholes is a program to compute the dynamics of a financial market containing derivative investment instruments. We test the variable \textit{xNPrimeofX} in the \textit{CNDF} function. According to the running logs, some of runs are crashed, and others generate corrupted results, none of which are correct.
Figure \ref{fig:res3} illustrates the percentage of crashed runs and completed (but with corrupted outputs) runs.
Due to the paper's focus (tool development), we will investigate the root cause of these crashes in the future. 

\subsubsection{XSBench}
\label{subsubsec:xsbench}

XS-Bench is a mini-app representing a key computational kernel of the Monte Carlo neutron transport algorithm. We test the variable \textit{a1} in the \textit{calculate\_macro\_xs} function. According to the standard output, although all injected programs can finish the execution, every program with fault injection either generates different output or runs lower with the same output, compared to the baseline program. Listing \ref{lst:res3} illustrates the different verification checksums generated by the baseline and injected programs. We note that the baseline programs cost about 127 seconds, but the programs with injected faults cost about 260 seconds.

\begin{lstlisting}[language=bash, caption={Difference between original \& injected programs.},label={lst:res3}]
$ diff llfi/baseline/golden_std_output \
> llfi/std_output/std_outputfile-run-0-0
46c46
< Verification checksum: 74966788162
---
> Verification checksum: 74966786750
\end{lstlisting}

\subsubsection{NPB-MG}
NPB-MG is a multi-grid (MG) method implemented in the NAS Parallel Benchmarks \cite{npbmg}. In numerical analysis, an MG method is an algorithm for solving differential equations using a hierarchy of discretizations. We test the variable \textit{a1} in the \textit{vranlc} function. We observe that the outputs of all the programs are corrupted with the tested fault types (including error distributions and error-bound modes). 

\subsection{Observation 1: Corrupted or Not? OR Slow Converge?}

According to Section \ref{subsec:res}, we observe that the programs with faults injected can crash (\textit{Black-Scholes}), generate incorrect results (\textit{HPCCG} and \textit{NPB-MG}), or take longer time to complete or converge (\textit{HPCCG} and \textit{XSBench}). In addition, some faults may have no impact on the program execution such as \textit{HPCCG}, which will be discussed in Section \ref{subsec:path}. 

Therefore, we can say that our tool can simulate different faulty scenarios and effectively guide users on how to use lossy compressors.  
As shown in Section \ref{subsubsec:hpcg}, we can find that, as the error bound increases, the $R_f$ becomes smaller, which means the final residual becomes larger; in other words, the program converges more slowly. This means that when users try to use lossy compression here, they have to be careful about the error bound to set. As shown in Section \ref{subsubsec:xsbench}, even if the simulation time becomes about twice longer, the program with injected fault still cannot get the correct output. This means that users cannot use lossy compression for this specific variable in \textit{XSbench}. 

\subsection{Observation 2: Execution Path Changed?}
\label{subsec:path}

\begin{figure}[H]
\vspace{-6mm}
\centering\small\sffamily
\begin{tikzpicture}
\node[draw, rounded corners=12pt, align=center, text width=5em] (00) {\texttt{Waxpby}\\\footnotesize(injected)};
\node[draw, rounded corners=12pt, align=center, text width=5em, below=of 00] (01) {\texttt{Waxpby}\\\footnotesize(with fault)};
\node[draw, rounded corners=12pt, align=center, text width=5em, right=of 00] (10) {\texttt{Waxpby}\\\footnotesize(injected)};
\node[fill=yellow!70, draw, rounded corners=12pt, align=center, text width=5em, below=of 10] (11) {\bfseries \texttt{Waxpby}\\\footnotesize(normal)};
\node[below=of 01] (02) {Normal Fault};
\node[below=of 11] (12) {Benign Fault};
\node[above=of 00] (col0) {};
\node[above=of 10] (col1) {};

\draw[-latex] (col0) -- (00);
\draw[-latex] (00) -- (01);
\draw[-latex] (01) -- (02);
\draw[-latex] (col1) -- (10);
\draw[-latex] (10) -- (11);
\draw[-latex] (11) -- (12);

\node[below=.15in of 10] (anil0) {};
\node[right=of anil0] (anil1) {\fontfamily{ugq}\selectfont\color{red}\footnotesize annihilation};
\draw[-latex] (anil1) -- (anil0);

\end{tikzpicture}
\vspace{-2mm}
\caption{Sample of Normal Fault and Benign Fault}
\label{fig:be}
\end{figure}
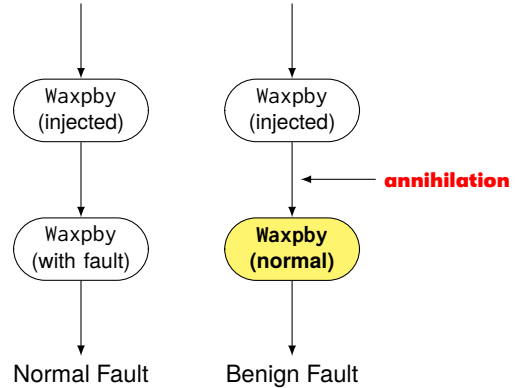

According to Table \ref{table:res1}, we observe that some injected faults do not have any noticeable impact on the program’s output. We call these faults \textit{Benign Faults}. Based on the trace file, we find that the fault was injected in the first loop but disappeared in the second loop. The first loop is located in line 5 of Listing \ref{lst:HPCCG}, and the second loop is located in line 7. We get such error propagation figures between benign fault and normal fault\footnote{\textit{Normal faults} are the faults having an impact on the program's final output.}, as shown Figure \ref{fig:be}. This demonstrates that users can use \textsc{LCFI} to effectively trace lossy compression error propagation.

\begin{lstlisting}[basicstyle=\ttfamily\scriptsize, caption={HPCCG benchmark in detail.},label={lst:HPCCG}]
  int print_freq = max_iter/10; 
  if (print_freq>50) print_freq=50;
  if (print_freq<1)  print_freq=1;
  // p is of length ncols, copy x to p for sparse MV ops
  TICK(); waxpby(nrow, 1.0, x, 0.0, x, p); TOCK(t2);
  TICK(); HPC_sparsemv(A, p, Ap); TOCK(t3);
  TICK(); waxpby(nrow, 1.0, b, -1.0, Ap, r); TOCK(t2);
  TICK(); ddot(nrow, r, r, &rtrans, t4); TOCK(t1);
  normr = sqrt(rtrans);
\end{lstlisting}

\section{Conclusion and Future Work}
\label{sec:conclusion}

In this paper, we propose and develop a new fault injector for lossy compression error called \textsc{LCFI} (\underline{L}ossy \underline{C}ompression \underline{F}ault \underline{I}njector). This tool can realize IR-level analysis for lossy compression errors. \textsc{LCFI} can provide useful insights for developers of lossy compression to design a better compression for specific HPC programs.
Based on our evaluation results, we find that different programs have different resilience on lossy compression errors. In specific programs, different variables or even the same variable in different locations may have different sensitivities to a given type of lossy compression error.
In the future, we plan to extend \textsc{LCFI} with OpenMP and GPU support, which will have broader prospects and applications.

\bibliographystyle{IEEEtran}
\bibliography{refs}

\end{document}